\documentclass[12pt]{article}\usepackage{quanfou}
\input{MyQcircuit.sty}
\begin{document}

\title{QuanFou, QuanGlue, QuanOracle and QuanShi,\\
Four Special Purpose Quantum Compilers }

\author{Robert R. Tucci\\
        P.O. Box 226\\
        Bedford,  MA   01730\\
        tucci@ar-tiste.com}

\date{ \today}

\maketitle

\vskip2cm
\section*{Abstract}

This paper introduces
QuanFou v1.1, QuanGlue v1.1, QuanOracle v1.1, QuanShi v1.1,
four Java applications
available for free. (Source code
included in the distribution.)
Each application
compiles a different
kind of
input quantum evolution operator.
The applications
output a quantum circuit
that equals the input
evolution operator.

\section{Introduction}

This paper introduces
QuanFou v1.1, QuanGlue v1.1, QuanOracle v1.1, QuanShi v1.1,
four Java applications
available\cite{QuanSuite} for free. (Source code
included in the distribution.)

In a previous paper\cite{qtree},
we introduced the Java applications
QuanTree and QuanLin.
These two applications
plus the four applications
introduced in this paper, are
part of a suite of Java
applications called QuanSuite.
QuanSuite applications
all depend heavily on a common class
library called QWalk. Each
QuanSuite application
compiles a different kind of
input quantum evolution operator.
The applications
output a quantum circuit
that equals the input
evolution operator.

Before reading this paper,
the reader should read
Ref.\cite{qtree}.
Many
explanations in
Ref.\cite{qtree}
still apply to this paper.
Rather than repeating
such explanations in this paper,
the reader will be
frequently referred to Ref.\cite{qtree}.

The input evolution operator $U$
for a QuanSuite application can
be specified either directly (e.g. in QuanFou, QuanShi),
or by giving a Hamiltonian $H$ such
that $U = e^{iH}$ (e.g. in QuanGlue and QuanOracle).

The standard definition of
the evolution operator
in Quantum Mechanics is
$U= e^{-itH}$, where
$t$ is time and $H$
is a Hamiltonian. Throughout
this paper, we will set
$t = -1$ so $U = e^{iH}$.
If $H$ is proportional
to a coupling constant $g$,
reference to time can be
 restored easily by
replacing the symbol $g$ by
$-tg$, and the symbol $H$ by $-tH$.

\section{QuanFou}

The input evolution operator for QuanFou
is $U_{DFT}$, the Discrete Fourier
Transform matrix, defined by:

\beq
(U_{DFT})_{p,q} = \frac{1}{\sqrt{\ns}}\omega^{pq}
\;\;{\rm where}\;\;\omega=e^{i\frac{2\pi}{\ns}}
\;,
\eeq
with $p,q\in Z_{0,\ns-1}$.

See Ref.\cite{Paulinesia}
for a review of how to compile $U_{DFT}$ exactly.

Since we use an exact (to numerical precision)
compilation of $U_{DFT}$, the Order of
the Suzuki(or other) Approximant
and the Number of Trots are two
parameters which do not arise in QuanFou
(unlike QuanTree and QuanLin).

\begin{figure}[h]
    \begin{center}
    \includegraphics[height=3.75in]{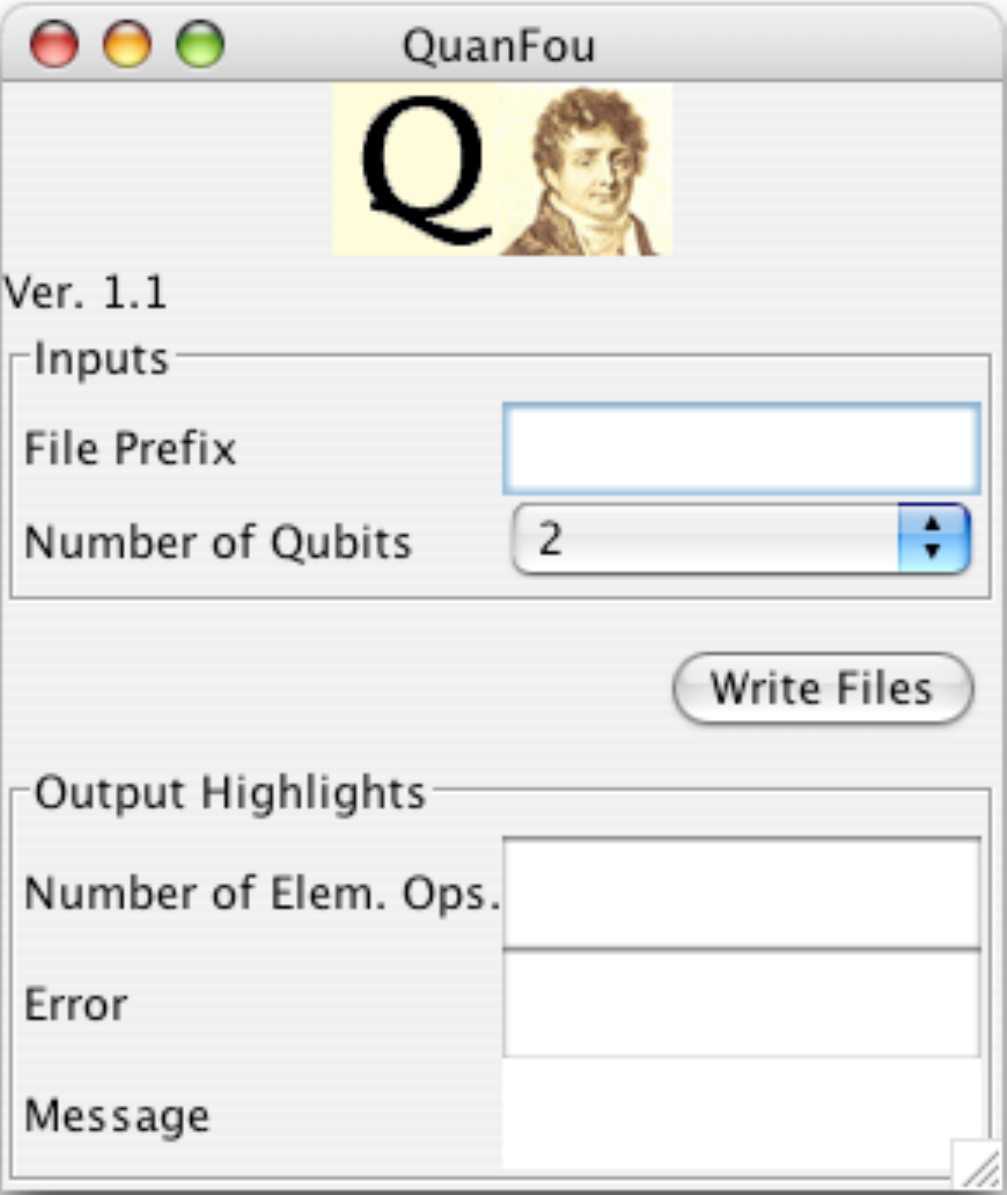}
    \caption{Control Panel of QuanFou}
    \label{fig-qfou-main}
    \end{center}
\end{figure}

Fig.\ref{fig-qfou-main} shows the
{\bf Control Panel} for QuanFou. This is the
main and only window of the application.

As to the input and output fields
in the Control Panel for QuanFou,
we've seen and explained these before in Ref.\cite{qtree}.

\begin{figure}[h]
\begin{center}
    \includegraphics[height=2.2in]{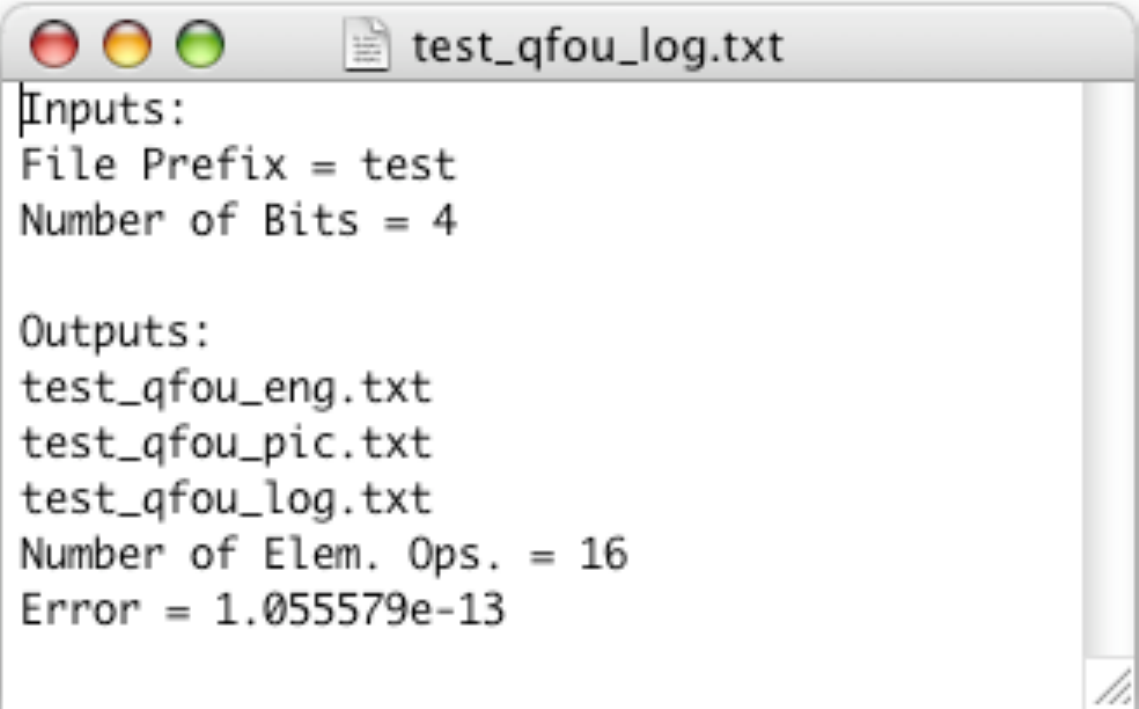}
    \caption{Log File generated by QuanFou}
    \label{fig-qfou-log}
\end{center}
\end{figure}

\begin{figure}[h]
\begin{center}
    \includegraphics[height=3in]{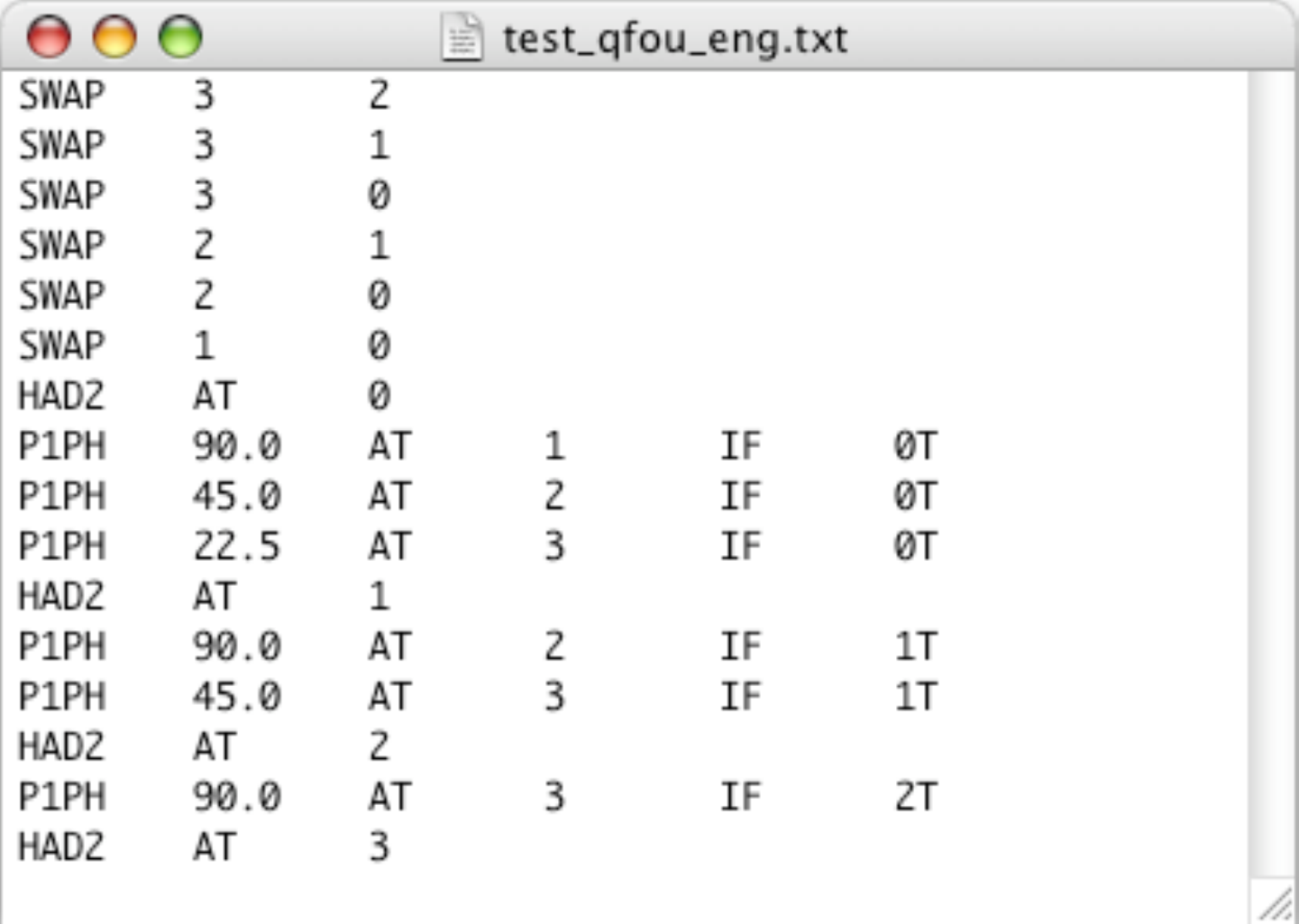}
    \caption{English File generated by QuanFou}
    \label{fig-qfou-eng}
\end{center}
\end{figure}

\begin{figure}[h]
\begin{center}
    \includegraphics[height=3in]{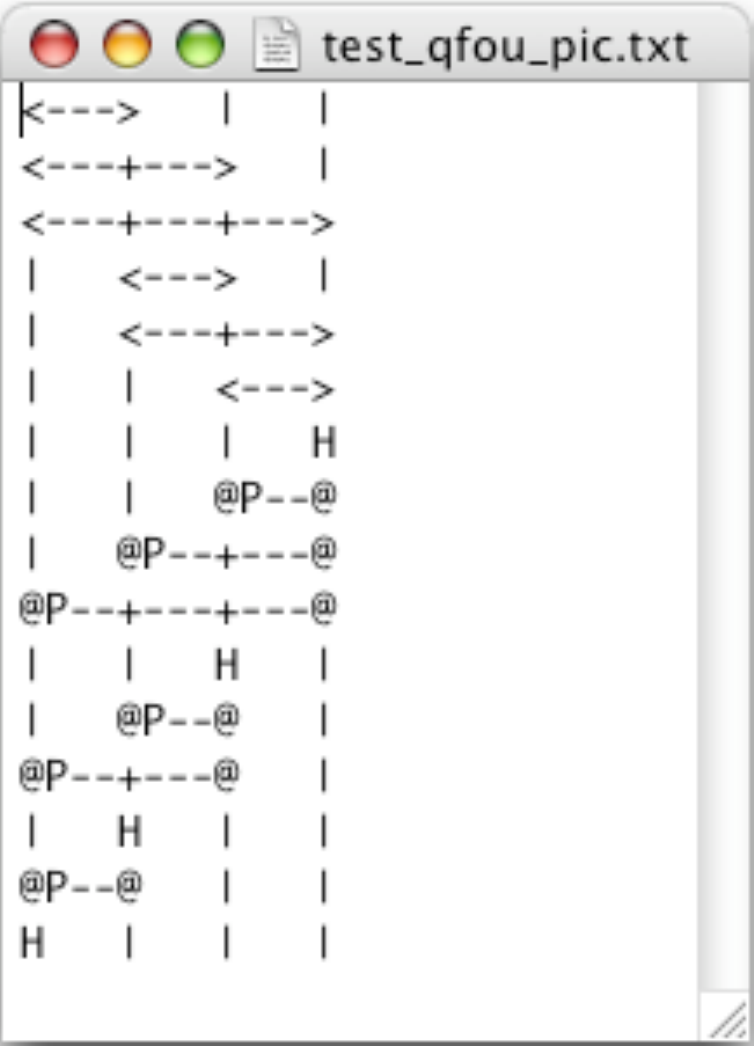}
    \caption{Picture File generated by QuanFou}
    \label{fig-qfou-pic}
\end{center}
\end{figure}

As to the output files (Log, English, Picture)
generated when we press the {\bf Write Files}
button, we've seen and explained these before in Ref.\cite{qtree}.
For example, Figs.\ref{fig-qfou-log},
\ref{fig-qfou-eng}, \ref{fig-qfou-pic}
show an instance of these
output files that was generated by
QuanFou.

\section{QuanGlue}

The input evolution operator for QuanGlue
is $U_{glue}=e^{iH_{glue}}$,
where:

\beq
H_{glue} = g(\ket{r_1}\bra{r_2} + h.c.)
\;,
\eeq
for some $r_1,r_2\in Z_{0,\ns-1}$.

Ref.\cite{Theory}
explains our method for compiling $U_{glue}$ exactly.

Since we use an exact (to numerical precision)
compilation of $U_{glue}$, the Order of
the Suzuki(or other) Approximant
and the Number of Trots are two
parameters which do not arise in QuanGlue
(unlike QuanTree and QuanLin).

\begin{figure}[h]
    \begin{center}
    \includegraphics[height=4.5in]{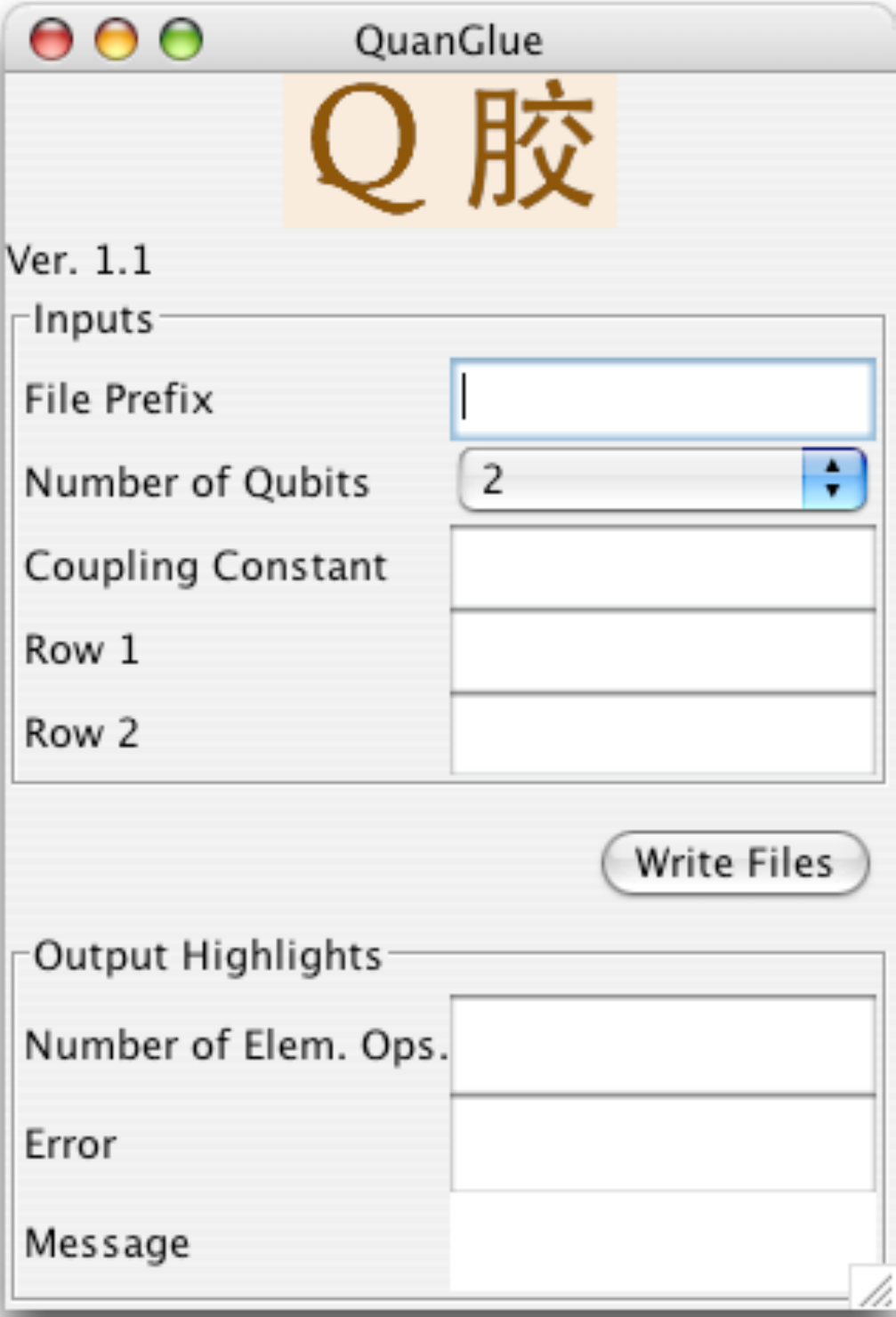}
    \caption{Control Panel of QuanGlue}
    \label{fig-qglue-main}
    \end{center}
\end{figure}

Fig.\ref{fig-qglue-main} shows the
{\bf Control Panel} for QuanGlue. This is the
main and only window of the application.

As to the input and output fields
in the Control Panel for QuanGlue,
we've seen and explained these before in Ref.\cite{qtree},
except for
the input fields {\bf Row 1} and
{\bf Row 2}.

\begin{description}
\item[Row 1, Row 2:]  Row 1 = $r_1$
and Row 2 = $r_2$ or vice versa,
where $r_1,r_2$ are the parameters defined
above, the two states being glued.
\end{description}

As to the output files (Log, English, Picture)
generated when we press the {\bf Write Files}
button, we've seen and explained these before in Ref.\cite{qtree}.

\section{QuanOracle}

Consider a tree with $N_{S,tree}$ states,
and
$\nlvs = \frac{N_{S,tree}}{2}$ leaves,
with leaf inputs $x_k \in Bool$
for $k\in Z_{0, \nlvs-1}$.
The input evolution operator for QuanOracle is
$U_{oracle} = e^{iH_{oracle}}$,
where

\beqa
H_{oracle}&=&
g
\left[
\begin{array}{cccc|cccc}
&&&&x_0&&&\\
&&&&&x_1&&\\
&&&&&&\ddots&\\
&&&&&&&x_{\nlvs-1}\\ \hline
x_0&&&&&&&\\
&x_1&&&&&&\\
&&\ddots&&&&&\\
&&&x_{\nlvs-1}&&&&
\end{array}
\right]
\;.
\eeqa

Ref.\cite{Theory}, in the appendix
for ``banded oracles",
explains our method for compiling $U_{oracle}$ exactly.

Since we use an exact (to numerical precision)
compilation of $U_{oracle}$, the Order of
the Suzuki(or other) Approximant
and the Number of Trots are two
parameters which do not arise in QuanOracle
(unlike QuanTree and QuanLin).

\begin{figure}[h]
    \begin{center}
    \includegraphics[height=4.5in]{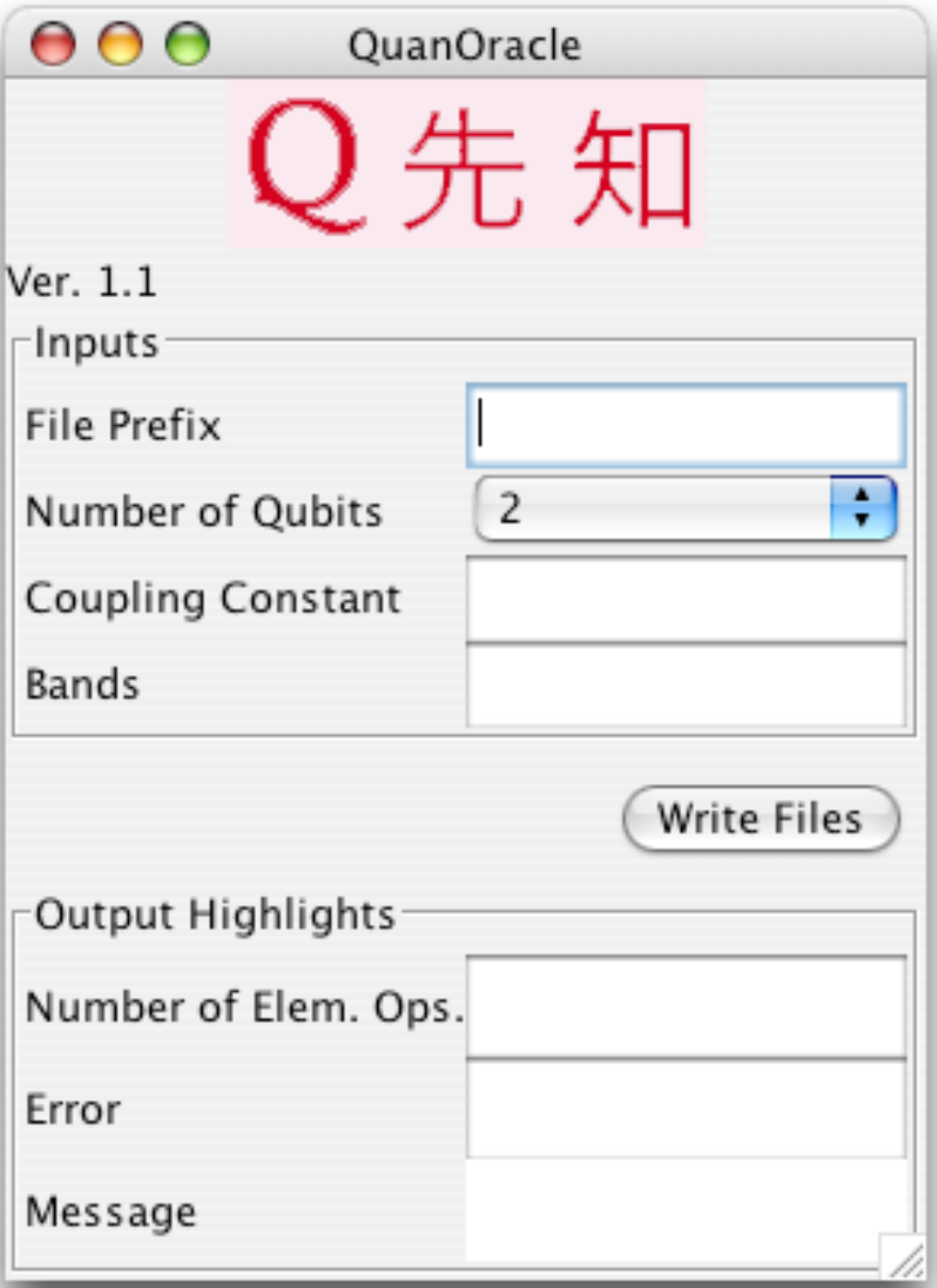}
    \caption{Control Panel of QuanOracle}
    \label{fig-qora-main}
    \end{center}
\end{figure}

Fig.\ref{fig-qora-main} shows the
{\bf Control Panel} for QuanOracle. This is the
main and only window of the application.

As to the input and output fields
in the Control Panel for QuanOracle,
we've seen and explained these before in Ref.\cite{qtree},
except for
the input field {\bf Bands}.

\begin{description}
\item[Bands:] You must enter here an even number
of integers separated by any non-integer,
non-white space symbols. Say you enter
$a_1,b_1,a_2,b_2,\ldots ,a_n,b_n$.
If $x_k\in Bool$ for $k\in Z_{0,\nlvs-1}$
are as defined above, then
$x_k=1$ iff $k\in
Z_{a_1,b_1}\cup Z_{a_2,b_2} \ldots \cup Z_{a_n,b_n}$.
Each set $Z_{a_k,b_k}$ is a ``band".
If $a_k=b_k$, the band has a single element.
QuanOracle checks that
$0\leq a_0$, $b_n\leq(\nlvs-1)$, and
$b_k-a_k\geq 0$ for all $k$.
It also checks that
$a_{k+1}-b_k\geq 2$. (If
$a_{k+1}-b_k= 1$, bands $k+1$ and $k$
can be merged. If
$a_{k+1}-b_k= 0, -1, -2, \ldots$,
bands $k+1$ and $k$ overlap.)
\end{description}

As to the output files (Log, English, Picture)
generated when we press the {\bf Write Files}
button, we've seen and explained these before in Ref.\cite{qtree}.

\section{QuanShi}

The input evolution operator for QuanShi
is the unitary operation $U_{shift}$ that takes:

\beq
\ket{x}\rarrow \ket{(x + t)\mod \ns}
\;,
\eeq
where
$x,t\in Z_{0,\ns-1}$, with
$\ns = 2^\nb$ for some positive integer $\nb$.
We call $t$ the {\bf state shift}.

$U_{shift}$ can be easily
expressed in matrix form.
For example, for $\ns=8$ and $t=3$,

\beq
U_{shift} =
\begin{array}{c||c|c|c|c|c|c|c|c|}
&\p{0}&\p{1}&\p{2}&\p{t}&&&& \\ \hline \hline
\p{0}&&&&1&&&& \\ \hline
\p{1}&&&&&1&&& \\ \hline
\p{2}&&&&&&1&& \\ \hline
\p{\vdots}&&&&&&&1& \\ \hline
&&&&&&&&1 \\ \hline
\p{\ns-t}&1&&&&&&& \\ \hline
&&1&&&&&& \\ \hline
\p{\ns-1}&&&1&&&&& \\ \hline
\end{array}
\;.
\eeq

Appendix \ref{app-shift}
explains our method for compiling $U_{shift}$ exactly.

Since we use an exact (to numerical precision)
compilation of $U_{shift}$, the Order of
the Suzuki(or other) Approximant
and the Number of Trots are two
parameters which do not arise in QuanShi
(unlike QuanTree and QuanLin).

\begin{figure}[h]
    \begin{center}
    \includegraphics[height=4in]{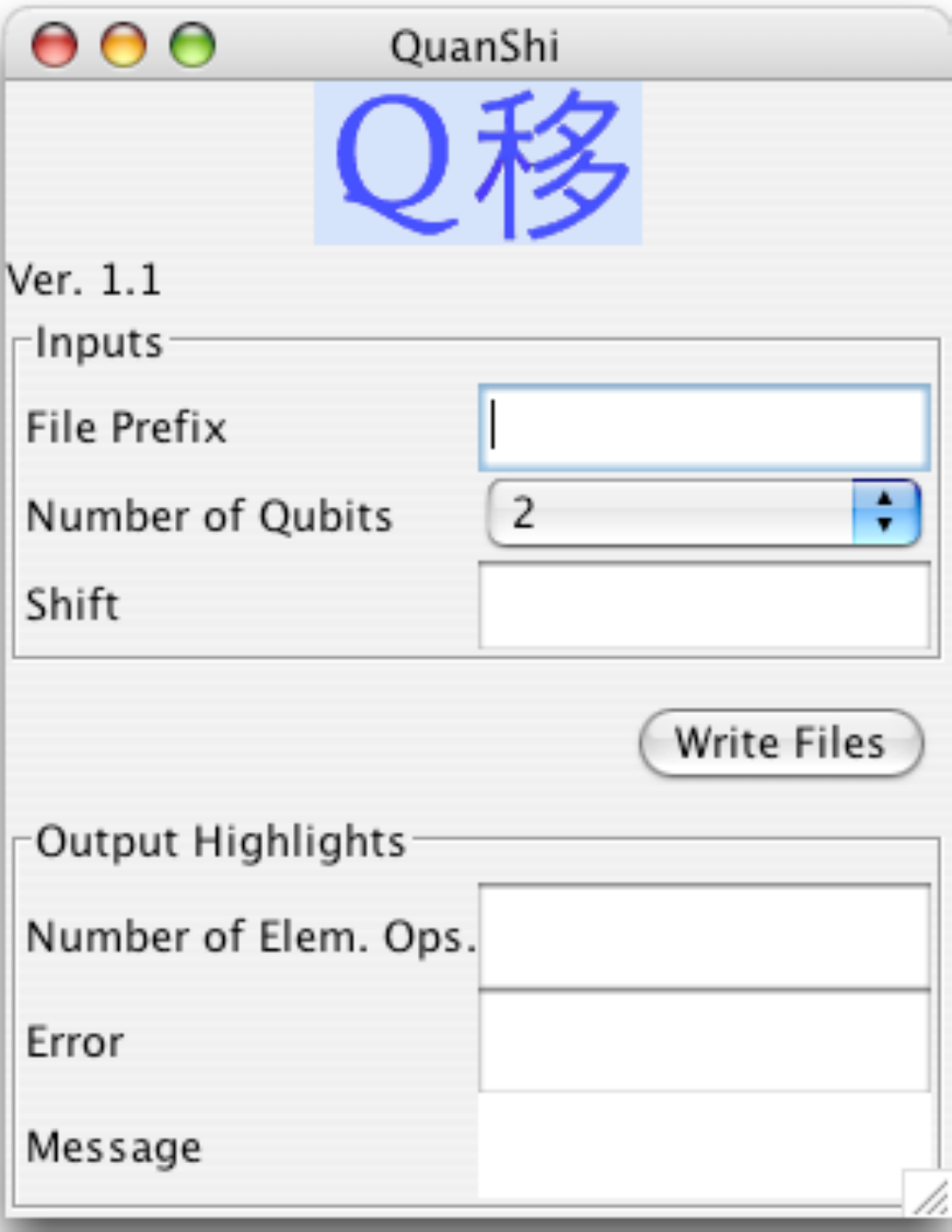}
    \caption{Control Panel of QuanShi}
    \label{fig-qshi-main}
    \end{center}
\end{figure}

Fig.\ref{fig-qshi-main} shows the
{\bf Control Panel} for QuanShi. This is the
main and only window of the application.

As to the input and output fields
in the Control Panel for QuanShi,
we've seen and explained these before in Ref.\cite{qtree},
except for
the input field {\bf Shift}.

\begin{description}
\item[Shift:] The parameter $t$ defined above.
QuanShi allows $-\ns < t < \ns$ and
interprets a shift by $-t$ as the inverse of
a shift by $t$.
\end{description}

As to the output files (Log, English, Picture)
generated when we press the {\bf Write Files}
button, we've seen and explained these before in Ref.\cite{qtree}.

\appendix
\section{Appendix: How to Compile a State Shift}
\label{app-shift}

In this appendix, we will show how to compile
the unitary operation $U_{shift}$ that takes

\beq
\ket{x}\rarrow \ket{(x + t)\mod \ns}
\;,
\eeq
where
$x,t\in Z_{0,\ns-1}$, with
$\ns = 2^\nb$ for some positive integer $\nb$.
We call $t$ the {\bf state shift}.

$U_{shift}$ can be easily
expressed in matrix form.
For example, for $\ns=8$ and $t=3$,

\beq
U_{shift} =
\begin{array}{c||c|c|c|c|c|c|c|c|}
&\p{0}&\p{1}&\p{2}&\p{t}&&&& \\ \hline \hline
\p{0}&&&&1&&&& \\ \hline
\p{1}&&&&&1&&& \\ \hline
\p{2}&&&&&&1&& \\ \hline
\p{\vdots}&&&&&&&1& \\ \hline
&&&&&&&&1 \\ \hline
\p{\ns-t}&1&&&&&&& \\ \hline
&&1&&&&&& \\ \hline
\p{\ns-1}&&&1&&&&& \\ \hline
\end{array}
\;.
\eeq

$U_{shift}$ is an example
of a circulant matrix. Ref.\cite{Theory}
reviews
the well known properties of circulant matrices.
Circulant matrices have a particularly
simple eigenvalue decomposition.
Define

\beq
\omega=e^{-i\frac{2\pi}{\ns}}
\;.
\eeq
Then, according to Ref.\cite{Theory},

\beq
U_{shift} = V D V^\dagger
\;,
\label{eq-ushi-decomp}
\eeq
where

\beq
V_{p,q} = \frac{1}{\sqrt{\ns}}\omega^{pq}
\;,
\eeq
and

\beq
D = diag(\lam_0, \lam_1,\ldots, \lam_{\ns-1})
\;\;{\rm with}\;\;
\lam_m = \omega^{mt}
\;.
\eeq
Note that $V^\dagger$ is the DFT matrix.

It follows that
\beq
D = \exp(-i\frac{2\pi}{\ns}t A)
\;,
\eeq
where

\beq
A =
\left[
\begin{array}{ccccc}
0&&&&\\
&1&&&\\
&&2&&\\
&&&\ddots&\\
&&&&\ns-1
\end{array}
\right]
=
\sum_{\vec{m}\in Bool^\nb}m P_{\vec{m}}
\;.
\eeq
We are using
$\vec{m} = (m_{\nb-1},\ldots,m_2,m_1,m_0)$,
and
$m = dec(\vec{m}) = 2^{\nb-1}m_{\nb-1}+
\cdots+ 2^2 m_2 + 2 m_1 + m_0$.

Consider $A$ for $\nb=2$:

\beqa
0P_{00} + 1P_{01} + 2P_{10} + 3P_{11}
&=&P_{01} + 2P_{1.} + P_{11}\\
&=&P_{.1} + 2P_{1.}\\
&=&n(0) + 2n(1)
\;.
\eeqa
Now consider $A$ for $\nb = 3$:

\beqa
\sum_{\vec{m}\in Bool^3}
(2^2 m_2 + 2 m_1 + m_0)P_{\vec{m}}&=&
\sum_{\vec{m}\in Bool^3}
(2^2 m_2P_{\vec{m}}) + 2n(1) + n(0)\\
&=&2^2P_{1..} + 2n(1) + n(0)\\
&=&2^2n(2) + 2n(1) + n(0)
\;.
\eeqa
This result can be easily
generalized using induction to
an arbitrary number of qubits.

An exact compilation of
$U_{shift}$ is now readily apparent
from Eq.(\ref{eq-ushi-decomp}).
The matrices $V$ and $V^\dagger$
are DFTs matrices so we know how to compile them.
The diagonal matrix  $D$  is also
easy to compile. For example, for
$\nb=3$,

\beq
D =\exp\left(-i\frac{2\pi}{\ns}t
[2^2n(2) + 2n(1) + n(0)]\right)=
\begin{array}{c}
\Qcircuit @C=1em @R=.25em @!R{
&\gate{e^{i\phi n}}&\qw
\\
&\gate{e^{i2\phi n}}&\qw
\\
&\gate{e^{i4\phi n}}&\qw
}
\end{array}
\;,
\eeq
where
$\phi = -\frac{2\pi}{\ns}t$.

\end{document}